\documentclass{PoS}

\title{Model-independent Measurement of the Atmospheric Muon
 Neutrino Energy Spectrum up to 2.5 PeV}

\ShortTitle{Model-independent Measurement of the Atmospheric Muon Neutrino Energy Spectrum}

\author{
The IceCube Collaboration\footnote{For collaboration list, see PoS(ICRC2019) 1177.}\\
{\itshape \href{http://icecube.wisc.edu/collaboration/authors/icrc19_icecube}{http://icecube.wisc.edu/collaboration/authors/icrc19\_icecube}}\\
E-mail: \email{jan.soedingrekso@icecube.wisc.edu, tobias.hoinka@icecube.wisc.edu, mathis.boerner@icecube.wisc.edu}
}

\abstract{
The IceCube Observatory at the South Pole allows the measurement of the diffuse neutrino flux. The assumption of specific flux parametrizations limits the range of spectral shapes. Given the increasing statistics of the data recorded, model-independent unfolding approaches can overcome these limitations. In this contribution, a model-independent approach to estimate the muon neutrino flux between 125 GeV and 2.5 PeV is presented. In order to extract muon neutrinos from the data taken by the detector, a machine-learning-based method is employed. This yields an efficient muon neutrino sample with a purity of over 99\,\%. The spectrum is estimated by a Likelihood-based unfolding technique involving a novel binning scheme using a decision tree on three years of IceCube data. This measurement provides the first model-independent muon neutrino spectrum for multiple years in this energy regime.\\
\vspace{4mm}
{\bfseries Corresponding author:}
\speaker{J. Soedingrekso}$^{1}$, T. Hoinka$^{1}$, M. B\"orner$^{1}$\\
{$^{1}$ \itshape TU Dortmund University}

}

\FullConference{36th International Cosmic Ray Conference -ICRC2019-\\
		July 24th - August 1st, 2019\\
		Madison, WI, U.S.A.}

\begin{document}

\section{Introduction}\label{sec:introduction}

IceCube is a cubic-kilometer neutrino detector installed in the ice at the geographic South Pole between depths of 1450\,m and 2450\,m, completed in 2010. Reconstruction of the direction, energy and flavor of the neutrinos relies on the optical detection of Cherenkov radiation emitted by charged particles produced in the interactions of neutrinos in the surrounding ice or the nearby bedrock.

Triggered IceCube events originate from multiple classes of particles and interactions, most of them being atmospheric muons, i.e. muons that are produced in air showers induced by charged cosmic rays. They account for an event rate of 3000\,Hz and thus for an overwhelming majority of all triggered events. Other classes of events detected by IceCube include muon neutrinos, however, they are not detected directly, instead, muons produced in charged current (CC) interactions with the detector matter are detected. As these muons lose their energy consistently over a long distance, their topology is referred to as \emph{track-like}. On the other hand, muon neutrinos can also interact in neutral current (NC) interactions that produce a hadronic shower and a muon neutrino. Events like that appear as a point-like energy loss and are called \emph{cascade-like} events. Electron neutrinos and tau neutrinos interacting in the detector volume can also produce events of the same event topology. This analysis focuses on track-like event topologies associated with muon neutrinos, which are also the most common type of neutrino interactions observed in the detector.

The observed flux of muon neutrinos has two distinct origins, cosmic ray interactions in the atmosphere and astrophysical fromo hadronic interactions at the source or during transit. The atmospheric flux is produced in cosmic ray interactions and subsequent meson decays in the upper atmosphere. Due to the mesons having a relatively long lifetime, the resulting flux of the \emph{conventional atmospheric neutrino flux} inherits a steeper spectral shape than the original cosmic-ray flux that is proportional to $E^{-3.7}$. Another expected, but no yet observed, component of the atmospheric neutrino flux, the \emph{prompt neutrino flux} is the product of the decay of charmed mesons which exhibit a much shorter lifetime, which limits the energy losses before the eventual decay into neutrinos. Thus, this component more closely follows the shape of the original cosmic-ray spectrum, which is proportional to $E^{-2.7}$.

The astrophysical neutrino flux, observed in a number of analyses \cite{HESE, HESE2}, is expected to have a spectral index of $\gamma=2.0$, which would appear as a significant flattening of the overall neutrino flux.

This analysis aims to perform a model-independent approach to estimating the spectrum of atmospheric muon neutrinos; instead of assuming a particular parametrization of the flux shape, an unfolding approach as presented in \cite{IC59} and \cite{IC79} was further improved and extended. In this analysis the flux in 14 energy bins from 125\,GeV to 2.5\,PeV is estimated using data recorded in three years of livetime.

\section{Event Selection}\label{sec:eventselection}

To extract a high-purity sample of muon neutrinos from the recorded data, machine learning techniques are used. The task corresponds to a binary classification problem, whereas neutrino-induced muon tracks are treated as the signal class and atmospheric muon tracks are treated as the background class.

All following steps of the classification are performed in a ten-fold cross validation scheme in order to quantify the performance of the classification procedure, including its stability.

Before the classification is performed, a set of high-level attributes needs to be selected from all available attributes in the dataset. This has two reasons: First, it dramatically speeds up the learning and application processes, and second, some of the attributes included in the dataset are well-known to exhibit disagreements between recorded data and simulations and should not be used to extract physical information from the dataset. Thus, the selection process has two steps: To quantify the disagreements between simulations and data, a random forest classifier is trained to distinguish between signal and background. Random forests offer the possibility to calculate the contribution of each attribute during the classification process, a quantity referred to as feature importance.

Using the feature importance, outliers, i.e. attributes that contribute disproportionally to the disagreements are flagged and excluded from the set of attributes. A more elaborate description of this procedure can be found in \cite{Brner2017MeasurementS}

To quantify the degree of disagreement between simulations and data, the ROC AUC, the area under the curve described by the receiver-operator characteristic, is calculated as a measure for the success of the aforementioned classification of simulations and data. The removal of the mismatched attributes reduced the AUC from $0.687 \pm 0.008$ to $0.624 \pm 0.006$.

The set of attributes used in this analysis is then further reduced by removing constant and strongly correlated and thus redundant attributes. To achieve that, of all attributes that have a Pearson correlation $\rho > 0.95$, only one is kept.

Following this pre-selection the attribute selection that aims to find a set of attributes that maximizes the relevance to the given classification task while minimizing the redundancy between the individual attributes. For this purpose the algorithm mRMR is used, which quantifies the relevance and redundancy using mutual information. \cite{DING2005}

\begin{figure}
    \centering
    \includegraphics[width=0.6\textwidth]{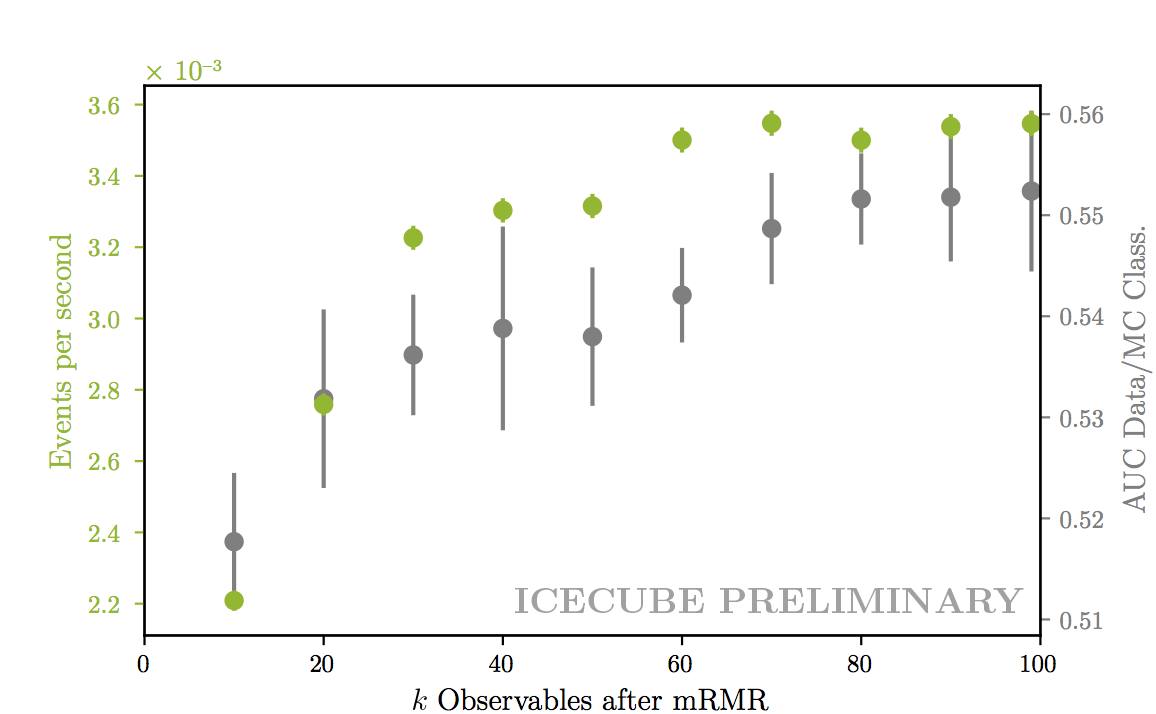}
    \caption{Performance of the event selection depending on the number of features $k$ selected for the analysis pipeline. In green is the resulting event rate of muon neutrinos, in gray the resulting AUC quantifying the amount of mismatches between MC and data.}
    \label{fig:nfeatures_perf}
\end{figure}
To decide how many features to keep, the neutrino rate and the AUC quantifying the mismatches are calculated. As can be seen in Fig.~\ref{fig:mc_unfolding}, the neutrino event rate plateaus at 60 attributes. Thus, a final number of 60 attributes are used for the classification, as increasing the number of attributes only increases the level of disagreements between the simulations and data.

\begin{figure}
    \begin{minipage}[b]{0.65\textwidth}
    \includegraphics[width=\textwidth]{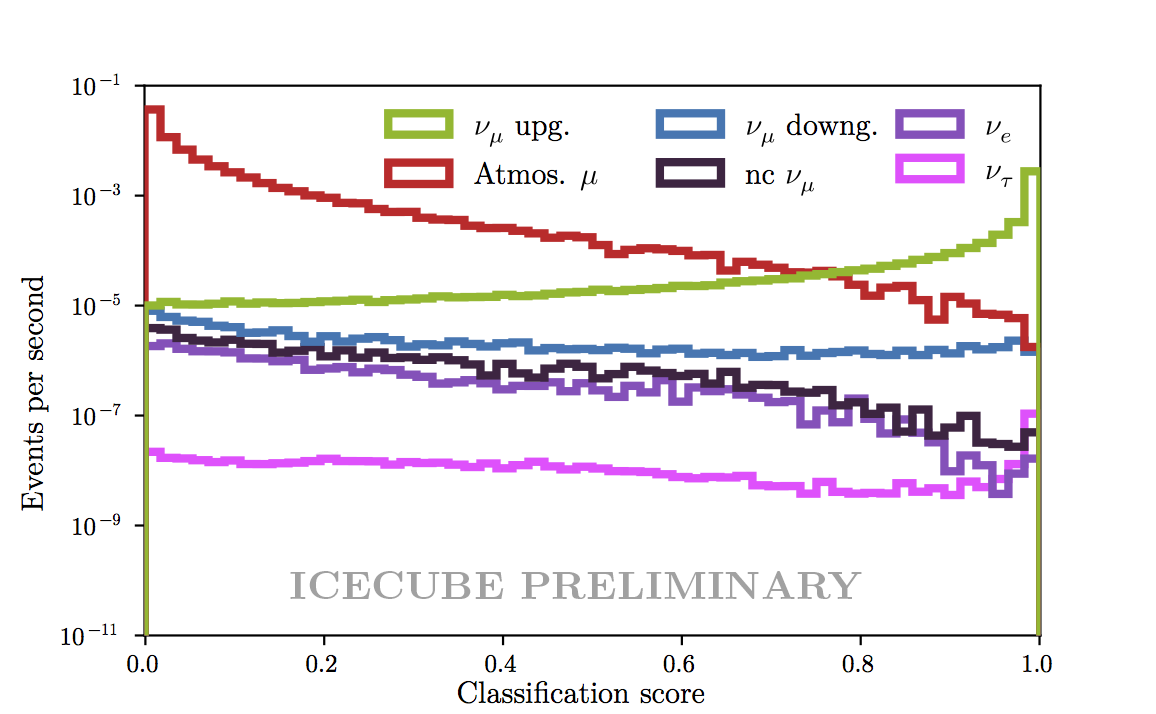}
    \end{minipage}
    \begin{minipage}[b]{0.34\textwidth}
    \includegraphics[width=\textwidth]{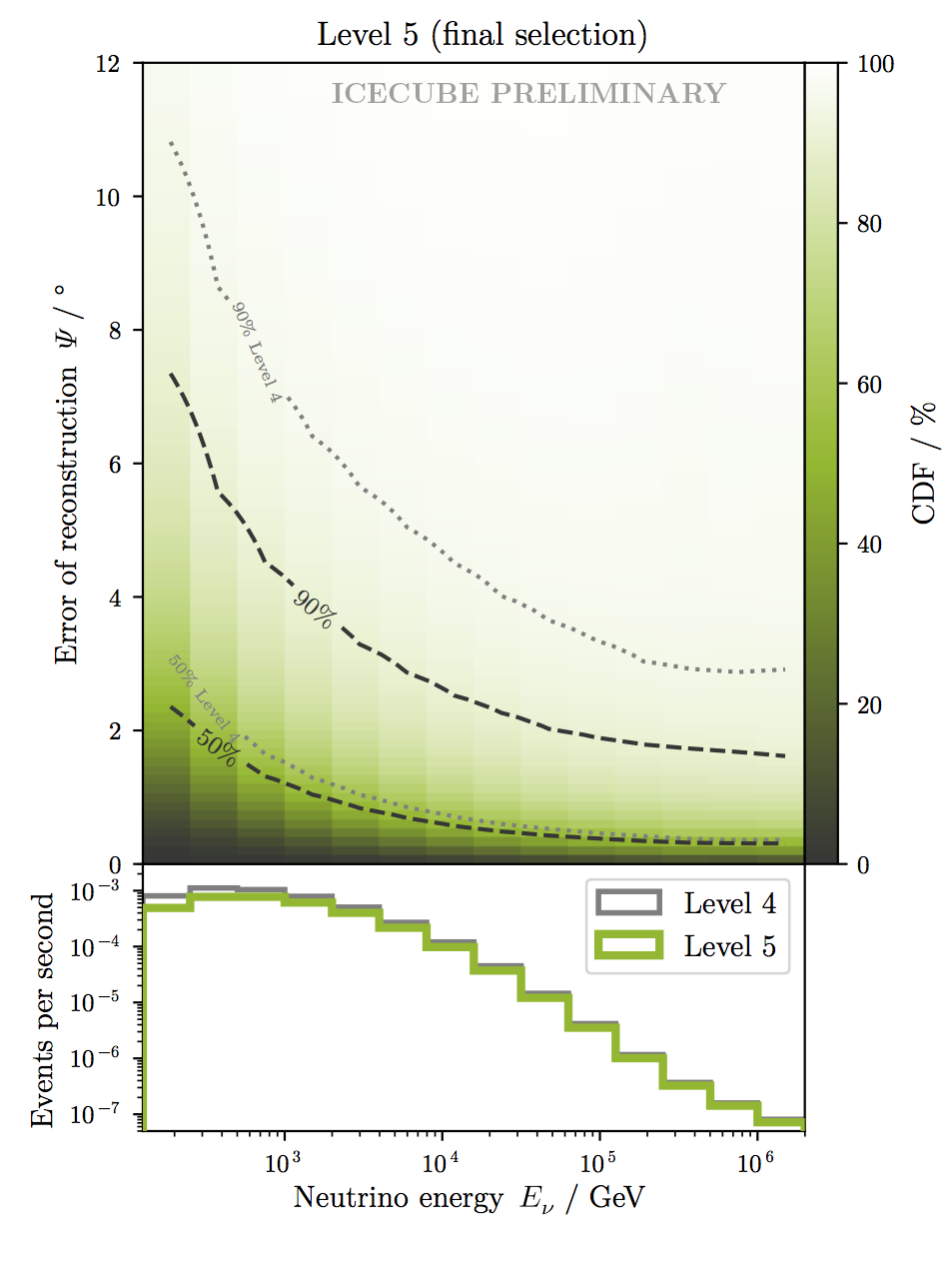}
    \end{minipage}
    \caption{\textit{Left:} Distribution of the classification scores for the different signal and background components of the simulation. \textit{Right:} Characterization of the final muon neutrino sample: In the top part the cumulative distribution of the reconstruction error for each energy bin is shown. The black dashed line shows the 50 and 90\,\% quantiles for the final event selection and the gray dashed line before the event selection. The bottom part shows the neutrino energy distribution before and after the event selection.}
    \label{fig:classif}
\end{figure}

After the attribute selection, a random forest classifier\cite{Breiman2001} is trained to differentiate between the signal class, i.e. upgoing with respect to the South Pole, well-reconstructed (meaning the angle between the reconstructed track and the true track is less than 5$^\circ$) and the background class, i.e. downgoing atmospheric muons. All other event classes are left out during the training. The distribution of the resulting random forest score for all event classes in the simulations can be seen in Fig.~\ref{fig:classif}. 

\section{Unfolding}\label{sec:unfolding}

Even though the energy of individual events can be reconstructed, there is a considerable amount of uncertainty involved with these estimations. These uncertainties result from two sources: on the one hand, the limited resolution of the detector, and on the other hand, an inability to know how long the neutrino-induced muon has traveled before reaching the detector volume. Additionally, there may be a background process present. The stochastic process that transfers the distribution of an unknown true quantity $x$ to the distribution of the measured observable $y$ can be described by the integral equation

\begin{equation}
g(y) = \int A(y | x)f(x) \mathrm{x} \mathrm{d} x + b(y)\,,\label{eq:fredholm}
\end{equation}

also referred to as the \emph{Fredholm Equation}. The function $A(y|x)$ is the response function and describes the conditional probability density of measuring the observable value $y$ given the underlying true quantity $x$. Using that function, the true distribution $f(x)$ can be translated into the distribution of the observable $g(y)$. The background $b(y)$ is added independently. In the case of this analysis however, the background is negligible due to the purity achieved in section x. 

The estimation of $f$ given $g$ is an inverse problem that requires finding a general representation of the functions $f$, $g$ and $A$ and solving the equation for $f$. The strategy used in this analysis firstly involves the discretization of both the observable and true quantity space in a number of bins, so that Eq.~\ref{eq:fredholm} translates into a much simpler linear algebra form,

\begin{equation}
\mathbf{g} = \mathbf{A}\mathbf{f}\,.
\end{equation}

Solving this equation is difficult due to the ill-conditioned matrix $\mathbf{A}$. To arrive at a plausible solution anyway, a Maximum likelihood approach is used that takes the Poissonian statistics of the vector $\mathbf{g}$ into account:

\begin{equation}
\mathcal{L}_\mathrm{Poisson}(\mathbf{g}|\mathbf{f}) = \prod_{i} \frac{(\mathbf{A}\mathbf{f})_i^{g_i}}{g_i!}\exp\left\{-(\mathbf{A}\mathbf{f})_i\right\}
\end{equation}

To counter the effects of the ill-conditioning of the problem, an additional prior is multiplied to the likelihood,
\begin{equation}
\mathcal{L}_\mathrm{reg}(\mathbf{f}) = \exp\left\{ -\frac{1}{2\tau} \log_{10}(\mathbf{A}_\mathrm{eff}^{-1}(\mathbf{f} + d \mathbf{1}))^\intercal \mathbf{C}^2 \log_{10}(\mathbf{A}_\mathrm{eff}^{-1} (\mathbf{f} + d \mathbf{1})) \right\}\,,
\end{equation}
whereas $\tau$ is the regularization strength, $\mathbf{A}_\mathrm{eff}$ is the effective area and $d$ is the regularization offset. The matrix $\mathbf{C}$ represents the second derivative and has the form

\begin{equation}
    \mathbf{C} =
    \left(
    \begin{array}{ccccccc}
         1 & -1 &  0 & \dots &&&\\
        -1 &  2 & -1 & \dots &&&\\
         0 & -1 &  2 & \dots &&&\\
         \vdots & \vdots & \vdots & \ddots & \vdots &\vdots &\vdots \\
         &&& \dots & 2 & -1 & 0\\
         &&& \dots & -1 & 2 & -1\\
         &&& \dots & 0 & -1 & 1\\
    \end{array}
    \right)
\end{equation}
Thus, the regularization term can be understood as a Gaussian prior that penalizes large second derivatives of the result vector $\mathbf{A}_\mathrm{eff}^{-1}\mathbf{f}$, which is a way of quantifying the smoothness of the result.

\begin{figure}
    \begin{minipage}[b]{0.6\textwidth}
    \includegraphics[width=\textwidth]{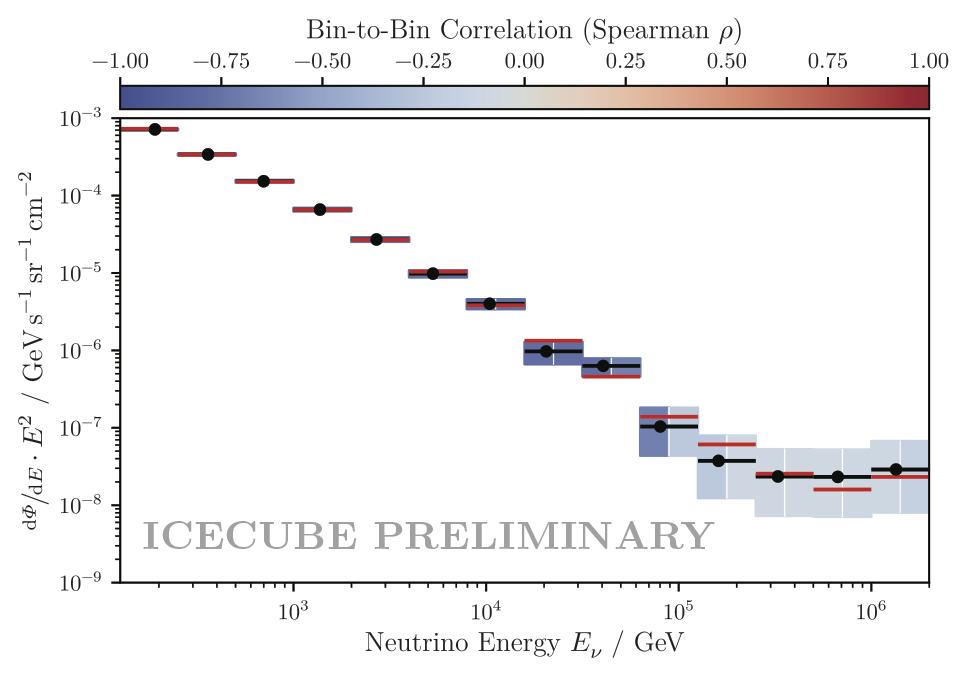}
    \end{minipage}
    \begin{minipage}[b]{0.4\textwidth}
    \includegraphics[width=\textwidth]{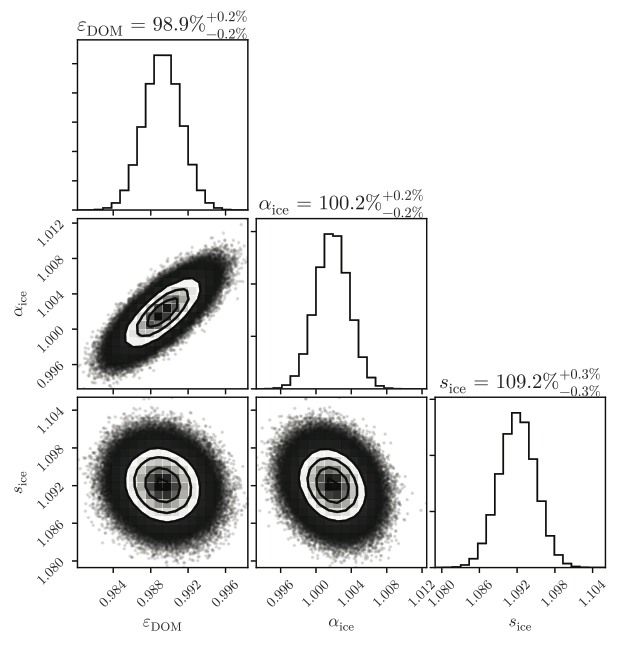}
    \end{minipage}
    \caption{Result of a dataset generated on the basis of simulation sampled from the model \cite{Aachen6yrs}. On the left, the unfolding result is shown; the red lines are the truth, the black lines the unfolding result. The height of the shaded areas indicate the marginalized errors, the color the correlation between neighbouring bins. On the right, the posterior distribution of the fitted systematics is shown.}
    \label{fig:mc_unfolding}
\end{figure}

This analysis includes several systematic parameters that aim to account for the uncertainties regarding the DOM efficiency $\epsilon_\mathrm{DOM}$, and the ice absorption and scattering coefficients $a_\mathrm{ice}$ and $s_\mathrm{ice}$. The effects of these systematics are included in the likelihood by parametrizing the effective area as a linear function of the systematic parameters $\mathbf{A}_\mathrm{eff}(\epsilon_\mathrm{DOM}, a_\mathrm{ice}, s_\mathrm{ice})$. This function is found by using different simulations that use systematic parameters that are slightly offset from the baseline.

The unfolding result is then found by maximizing the likelihood function
\begin{equation}
    \hat \mathbf{f} = \mathrm{argmax}(\mathcal{L}_\mathrm{Poisson}(\mathbf{g}|\mathbf{f}) \cdot \mathcal{L}_\mathrm{reg}(\mathbf{f}))\,.
\end{equation}
The resulting posterior distribution is then sampled using a Markov Chain Monte Carlo scheme. From this sample of the posterior pdf, a quasi-$p$-value
\begin{equation}
    p'(\mathbf{f}_\mathrm{test}) = \frac{1}{N_\mathrm{MCMC}}\sum_{i=1}^{N_\mathrm{MCMC}}
    \left\{\begin{array}{ll}
        1 & \quad p(\mathbf{f}_\mathrm{test}|\mathbf{g}) < p(\mathbf{f}_i|\mathbf{g})\\
        0 & \quad p(\mathbf{f}_\mathrm{test}|\mathbf{g}) \ge p(\mathbf{f}_i|\mathbf{g})
    \end{array}\right.\label{eq:ts}
\end{equation}
can be calculated, where $N_\mathrm{MCMC}$ is the number of samples drawn in the MCMC scheme, $p(\mathbf{f}_\mathrm{test}|\mathbf{g})$ the posterior pdf of the injected truth and $p(\mathbf{f}_i|\mathbf{g})$ the posterior pdf of the $i$-th sample of the MCMC. In the case of an unbiased estimation, the true value $\mathbf{f}_\mathrm{true}$ obeys the posterior distribution as given by the MCMC sample. Thus, ideally, the fraction of samples in the MCMC that have a smaller posterior pdf than the truth should be uniformly distributed in $[0,1]$.

\begin{figure}
    \centering
    \includegraphics{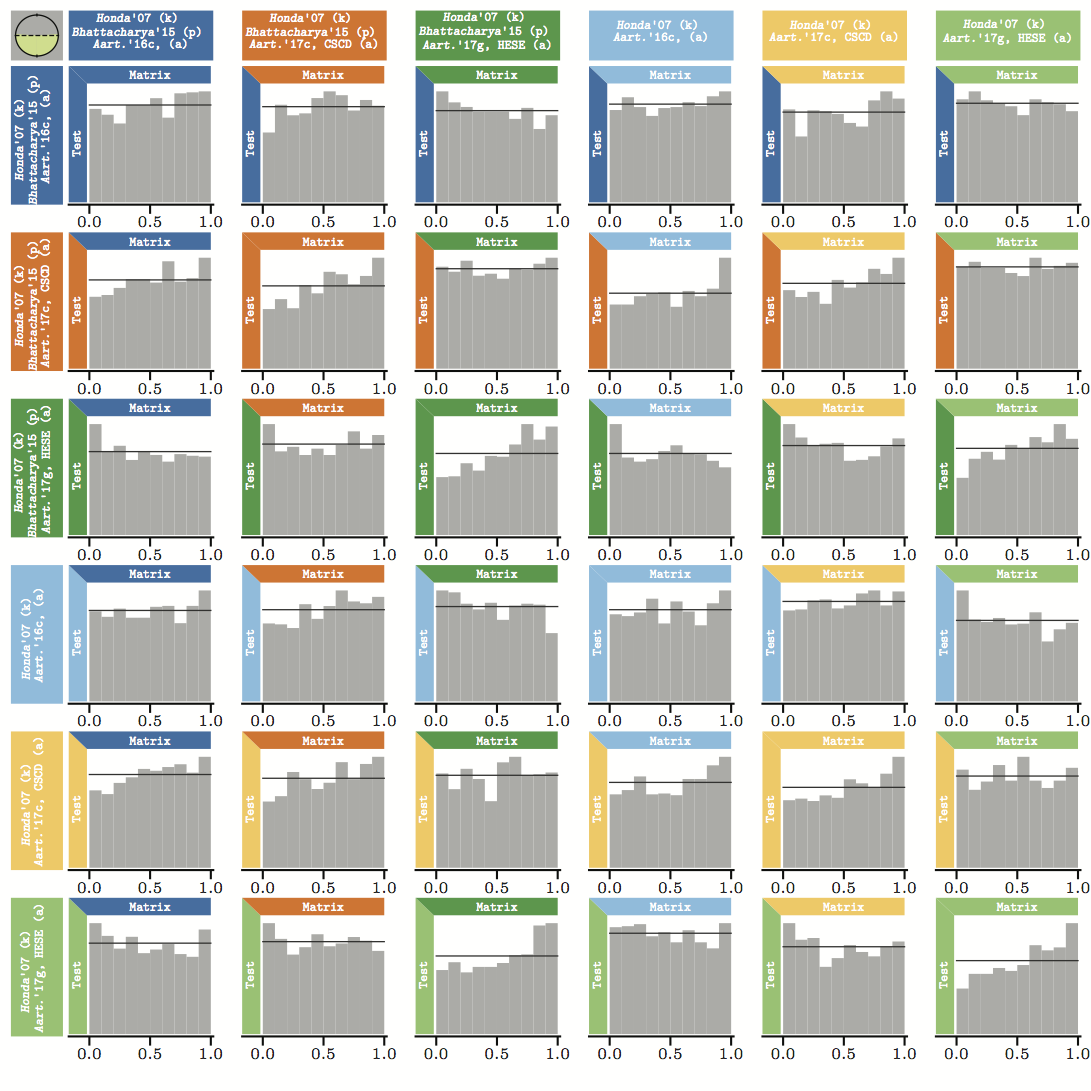}
    \caption{Distributions of the quasi-$p$-value defined in Eq.~\ref{eq:ts} for different combinations of spectra for the construction of the transfer matrix $\mathbf{A}$ and the unfolded spectra. For the spectra, a conventional flux from \cite{Honda2007} is assumed. The six different combinations of fluxes are the conventional flux added to an astrophysical flux from \cite{ICRC17hans} (orange and yellow), an astrophysical flux from \cite{ICRC17HESE} (green and light green), an astrophysical flux from \cite{Aachen6yrs} (blue and light blue) and each of these with a prompt component from \cite{Bhattacharya2015} (light green, light blue and yellow).}
    \label{fig:pval_dist}
\end{figure}

The free parameters of the regularization term, $\tau$ and $d$ are found by performing the unfolding scheme for several artificial datasets derived from re-weighted simulations using different flux models. The bias of these unfoldings is then checked and a set of regularization parameters is chosen that is unbiased for reasonable flux models. Fig.~\ref{fig:mc_unfolding} shows an example for such an unfolding. The unfolding scheme aims to unfold the flux values in 14 energy bins. Additionally to the unfolded flux values, the fit recovers the three systematic parameters considered in the likelihood. Fig~\ref{fig:pval_dist} shows the distributions of Eq.~\ref{eq:ts} for multiple combinations of flux models to build the migration matrix $\mathrm{A}$ and the unfolded spectra $\mathbf{f}_\mathrm{test}$. For each combination of flux models, 1000 tests are performed containing 500,000 MCMC samples each. The distributions are in fair compatibility with uniformity and thus show that for reasonable assumed flux models, the unfolding procedure as presented in this work exhibits low biases. This confirms the ability of this procedure to unfold the atmospheric neutrino flux largely independent of the model, i.e. systematic flux uncertainties.

\bibliographystyle{ICRC}
\bibliography{references}

\end{document}